\documentclass{article}
\usepackage{amsmath, amssymb, amsfonts}
\title{ Accelerating imperfect fluid } 
\author{Hristu Culetu, \\Ovidius University, Department of Physics and Electronics, \\ Bld. Mamaia  124, 900527 Constanta, Romania, \\e-mail : hculetu@yahoo.com}

\begin{document}
\numberwithin{equation}{section}
\pagenumbering{arabic}
\maketitle
\newcommand{\fv}{\boldsymbol{f}}
\newcommand{\tv}{\boldsymbol{t}}
\newcommand{\gv}{\boldsymbol{g}}
\newcommand{\OV}{\boldsymbol{O}}
\newcommand{\wv}{\boldsymbol{w}}
\newcommand{\WV}{\boldsymbol{W}}
\newcommand{\NV}{\boldsymbol{N}}
\newcommand{\hv}{\boldsymbol{h}}
\newcommand{\yv}{\boldsymbol{y}}
\newcommand{\RE}{\textrm{Re}}
\newcommand{\IM}{\textrm{Im}}
\newcommand{\rot}{\textrm{rot}}
\newcommand{\dv}{\boldsymbol{d}}
\newcommand{\grad}{\textrm{grad}}
\newcommand{\Tr}{\textrm{Tr}}
\newcommand{\ua}{\uparrow}
\newcommand{\da}{\downarrow}
\newcommand{\ct}{\textrm{const}}
\newcommand{\xv}{\boldsymbol{x}}
\newcommand{\mv}{\boldsymbol{m}}
\newcommand{\rv}{\boldsymbol{r}}
\newcommand{\kv}{\boldsymbol{k}}
\newcommand{\VE}{\boldsymbol{V}}
\newcommand{\sv}{\boldsymbol{s}}
\newcommand{\RV}{\boldsymbol{R}}
\newcommand{\pv}{\boldsymbol{p}}
\newcommand{\PV}{\boldsymbol{P}}
\newcommand{\EV}{\boldsymbol{E}}
\newcommand{\DV}{\boldsymbol{D}}
\newcommand{\BV}{\boldsymbol{B}}
\newcommand{\HV}{\boldsymbol{H}}
\newcommand{\MV}{\boldsymbol{M}}
\newcommand{\be}{\begin{equation}}
\newcommand{\ee}{\end{equation}}
\newcommand{\ba}{\begin{eqnarray}}
\newcommand{\ea}{\end{eqnarray}}
\newcommand{\bq}{\begin{eqnarray*}}
\newcommand{\eq}{\end{eqnarray*}}
\newcommand{\pa}{\partial}
\newcommand{\f}{\frac}
\newcommand{\FV}{\boldsymbol{F}}
\newcommand{\ve}{\boldsymbol{v}}
\newcommand{\AV}{\boldsymbol{A}}
\newcommand{\jv}{\boldsymbol{j}}
\newcommand{\LV}{\boldsymbol{L}}
\newcommand{\SV}{\boldsymbol{S}}
\newcommand{\av}{\boldsymbol{a}}
\newcommand{\qv}{\boldsymbol{q}}
\newcommand{\QV}{\boldsymbol{Q}}
\newcommand{\ev}{\boldsymbol{e}}
\newcommand{\uv}{\boldsymbol{u}}
\newcommand{\KV}{\boldsymbol{K}}
\newcommand{\ro}{\boldsymbol{\rho}}
\newcommand{\si}{\boldsymbol{\sigma}}
\newcommand{\thv}{\boldsymbol{\theta}}
\newcommand{\bv}{\boldsymbol{b}}
\newcommand{\JV}{\boldsymbol{J}}
\newcommand{\nv}{\boldsymbol{n}}
\newcommand{\lv}{\boldsymbol{l}}
\newcommand{\om}{\boldsymbol{\omega}}
\newcommand{\Om}{\boldsymbol{\Omega}}
\newcommand{\Piv}{\boldsymbol{\Pi}}
\newcommand{\UV}{\boldsymbol{U}}
\newcommand{\iv}{\boldsymbol{i}}
\newcommand{\nuv}{\boldsymbol{\nu}}
\newcommand{\muv}{\boldsymbol{\mu}}
\newcommand{\lm}{\boldsymbol{\lambda}}
\newcommand{\Lm}{\boldsymbol{\Lambda}}
\newcommand{\opsi}{\overline{\psi}}
\renewcommand{\tan}{\textrm{tg}}
\renewcommand{\cot}{\textrm{ctg}}
\renewcommand{\sinh}{\textrm{sh}}
\renewcommand{\cosh}{\textrm{ch}}
\renewcommand{\tanh}{\textrm{th}}
\renewcommand{\coth}{\textrm{cth}}

\begin{abstract}
An inhomogeneous fluid in accelerated motion is investigated. When the velocity field $v(x)$ is not constant, the geometry viewed by a static observer is curved, as if the observer were immersed in a gravitational field. A velocity-dependent semiclassical gravitational potential is introduced, which obeys an Yukawa-type equation, written in Cartesian coordinates. The timelike and null geodesic equations are investigated. One finds that the fluid has zero energy density corresponding to the perfect fluid part but nonzero anisotropic energy density. The pressures will no longer depend on $\hbar$ for time intervals $t>>1/m$, where $m$ is the field mass.
 \end{abstract}

 \section{Introduction}
Analogue spacetimes provide general relativists with concrete physical models to help focus their approach. Conversely, the techniques of curved spacetime may help improve our understanding of condensed matter physics \cite{LSTV, MV, BLV, IC, JS, MN}. The first analogue spacetime with optics-based contribution was introduced by Gordon - the so-called Gordon metric \cite{WG}, but a simple model for a dumb (''acoustic'') black hole (BH) has been introduced by Unruh \cite{WU}, where sound is dragged along by a moving fluid. 

The key idea in analogue spaces is to consider some sort of excitation of some kind of background \cite{MV}. Visser \cite{MV2} showed that the connection between the Schwarzschild geometry and the acoustic metric is best expressed by the Schwarzschild line-element in Painleve-Gullstrand (PG) coordinates \cite{KW, HC, TWZ}. He studied the propagation of sound waves (the high frequency limit) through a moving inhomogeneous medium. The fluid velocity plays the role of the velocity $\sqrt{2M/r}$ from the PG metric, where $M$ is the BH mass. 
 In \cite{MV} he also investigated in detail the optical Gordon metric and expressed the opinion that there is not very useful to write down the Einstein equations for this optical metric. To generalize Gordon's metric, Visser takes into account a 4-velocity which is both space and time dependent. 

Using a somewhat different approach, we attempt to calculate the stress tensor and the curvature invariants for the curved spacetime created by a moving fluid, say along the x-axis, with a velocity vector field $v(x)$. We also stress that the geometry is formally equivalent to an acoustic metric, even though the velocity $c$ from the line-element is the velocity of light in vacuo, and not the velocity of sound in some condensed matter system. 

The letter is motivated by the special case we encounter by replacing the velocity of sound in the acoustic metric with the velocity of light. In contrast to other velocities, it is Lorentz-invariant, according to the second principle of the Special Relativity. As we shell see, the motion of the fluid is determined by the velocity potential $\Phi (x)$, proportional to $v^{2}(x)$. Compared to other authors, $v(x)$ depends on Planck's constant and vanishes when $\hbar \rightarrow 0$, so that our approach is semiclassical. 

 The main goal of the letter is to study the acoustic line-element from the viewpoint of General Relativity, by searching the sources where the curvature is originating from, namely studying the properties of the energy-momentum tensor, expressed in terms of $v(x)$ and its derivatives. We also looked for the reasons why the fluid stress tensor has only transversal nonzero components, with the fluid chosen to flow on the x-direction. We found that it has nonzero shear viscosity - a feature that may generate anisotropic stresses \cite{BM}. Following Barrow and Maartens \cite{BM}, we introduced an anisotropic energy density $\rho_{c}$ (the ''isotropic'' energy density is vanishing). We also investigated the structure of the energy-momentum tensor $T^{a}_{~b}$, compared it with the stress tensor on a hypersurface of constant potential $\Phi(x)$ (or constant speed $v(x)$) \cite{FM} and pointed that they are very similar.  

 Another purpose was to find the timelike and null geodesics of test particles in the acoustic geometry, for some particular value of $v(x)$, when the potential $\Phi (x)$ satisfies an Yukawa-type equation.

We introduce in Sec.2 the geometry viewed by an observer w.r.t. whom the medium has velocity $v(x)$ and exhibit its basic quantities from the point of view of General Relativity (GR). We also study the particular situation when the stress tensor vanishes for some form of the velocity field, when the metric is of Rindler-type. Sec.3 is devoted to the properties of the imperfect fluid, its anisotropic stresses and the similarities between the energy momentum tensor of the fluid and the assumed surface energy density tensor on a hypersurface of constant velocity.. In Sec.4 a scalar gravitational potential is introduced and one investigates its relation with the de Broglie-Bohm quantum potential; in addition, the geodesics in a geometry where the velocity field depends on that scalar potential are calculated. The paper ends with few conclusions in Sec.5.

We use geometrical units (c = G = $\hbar$ = 1) throughout the paper, unless otherwise specified.

\section{Moving fluid geometry}
Let us consider an inhomogeneous fluid \cite{MV2} with the velocity $v(x)$, flowing along the x-axis of the Cartesian coordinates $(t, x, y, z)$ w.r.t. a static observer
    \begin{equation}
  ds^{2} = -(1- v^{2}(x)) dt^{2} - 2v(x)dt dx + dx^{2} + dy^{2} + dz^{2}, ~~~~v(x) \geq 0.
 \label{2.1}
 \end{equation}
The cases $v = 0$ or $v = const.$ give us the Minkowski flat space. However, a variable velocity leads to a curved metric, with a nonzero source $T_{ab}$ in Einstein's equation $G_{ab} = 8\pi T_{ab}$ ($a,b$ run from $t$ to $z$). The geometry (2.1) is a solution of the Einstein equations provided
    \begin{equation}
		8\pi T^{y}_{~y} = 8\pi T^{z}_{~z} = -v'^{2} - vv'' = -\frac{d^{2}}{dx^{2}}\left(\frac{v^{2}}{2}\right), 
 \label{2.2}
 \end{equation}
with $v' = dv(x)/dx$ and all the other components of the stress tensor are vanishing. Therefore, the fluid is imperfect \cite{FC} (no $T^{x}_{~x}$ stresses). The scalar curvature of (2.1) is $R^{a}_{~a} = 2(v'^{2} + vv'')$ and the Kretschmann scalar $K = (R^{a}_{~a})^{2}$. 

We look now for the expression of $v(x)$ which obeys the condition (vacuum solution)
    \begin{equation}
		 v'^{2} + vv'' = 0
 \label{2.3}
 \end{equation}
With appropriate initial conditions, (2.3) yields
    \begin{equation}
		v(x) = \sqrt{2gx},
 \label{2.4}
 \end{equation}
where $g$ is a positive constant and $x$ is taken to be nonnegative. In addition, with $v(x)$ given by (2.4), one finds that the Riemann tensor vanishes so that the metric (2.1) is flat. We show now this is nothing but the Rindler spacetime, written in different coordinates. With $v(x)$ from (2.4), eq. (2.1) appears as
    \begin{equation}
  ds^{2} = -(1- 2gx) dt^{2} - 2\sqrt{2gx}dt dx + dx^{2} + dy^{2} + dz^{2}.
 \label{2.5}
 \end{equation}
To get rid of the off-diagonal term, we change to a new time $u$
    \begin{equation}
		du = dt + \frac{\sqrt{2gx}}{1-2gx}dx.
 \label{2.6}
 \end{equation}
When one introduces (2.6) in (2.5), we get
    \begin{equation}
	  ds^{2} = -(1- 2gx) du^{2} + \frac{dx^{2}}{1- 2gx} + dy^{2} + dz^{2},	
 \label{2.7}
 \end{equation}
that is the Rindler metric, with $g$ the constant acceleration. In other words, the variable $v(x)$ from (2.4) is the only expression leading to a flat geometry.

If we name $-\frac{v^{2}(x)}{2} = \Phi (x)$, one may consider $\Phi (x)$ as a velocity potential. In that case, constant $v$ (inertial motion) means constant scalar potential $\Phi$. When $v(x)$ is given by (2.4), we have $\Phi (x) = -gx$ and $-d\Phi /dx$ gives the constant acceleration $g$. Hence, we may consider $\Phi (x)$ as a gravitational potential and so one could define the inertial motion as a motion in a constant gravitational potential \cite{HC2}. 

We take the velocity 4-vector of an observer comoving with the medium to be given by 
    \begin{equation}
		u^{a} = (1, v(x), 0, 0),~~~u^{a}u_{a} = -1.
 \label{2.8}
 \end{equation}
One easily finds that $a^{b} = u^{a} \nabla_{a}u^{b} = 0$; that is, $u^{a}$ is tangent to the timelike geodesics. It is worth noting from (2.8) that the time $t$ is the proper time $\tau$ of the comoving observer, so that $u^{x} = dx/d\tau = dx/dt = v(x) = \sqrt{2gx}$, whence $x(t) = gt^{2}/2$ and $v(x(t)) = gt$, as in Newtonian mechanics.

\section{Anisotropic stresses}
As the fluid given by (2.2) is imperfect, we write down the general expression of $T_{ab}$ in terms of the anisotropic tensor $\pi _{ab}$, for any $v(x)$. We have \cite{ZVA}
  \begin{equation}
	 T_{ab} = \rho u_{a} u_{b} + p h_{ab} + \pi_{ab} +  u_{a} q_{b} + u_{b} q_{a},
 \label{3.1}
 \end{equation}
where the energy density $\rho =  T_{ab}u^{a} u^{b}$, the isotropic pressure $p = (1/3) h^{ab} T_{ab}$, the energy flux $q_{a} = -h_{ab}  T^{b}_{c}u^{c}$ and the anisotropic tensor 
  \begin{equation}
	\pi_{ab} = \frac{1}{2} \left( h_{ac} h_{bd}T^{cd} + h_{bc} h_{ad}T^{cd}\right) -\frac{1}{3} h_{ab} h^{cd}T_{cd},
 \label{3.2}
 \end{equation}
with $h_{ab} = g_{ab} + u_{a} u_{b}$ and $h_{ab} u^{b} = \pi_{ab} u^{b} = u^{b} q_{b} = \pi_{b}^{~b} = 0$. The components of $T^{a}_{~b}$ from (2.2) and of $u^{a}$ from (2.8) yield
  \begin{equation}
	\rho = 0,~~~p = \frac{2}{3} T^{y}_{~y} = \frac{2}{3} T^{z}_{~z} = -\frac{1}{12\pi} (v'^{2} + vv''),~~~q^{a} = 0.
 \label{3.3}
 \end{equation}
As far as the anisotropic stresses are concerned, the only nonzero components are
  \begin{equation}
	\pi^{x}_{~x} = -2 \pi^{y}_{~y} = -2 \pi^{z}_{~z} = -\frac{\pi^{x}_{~t}}{v} = \frac{1}{12\pi}(v'^{2} + vv''),~~~ \pi^{b}_{~b} = 0.
 \label{3.4}
 \end{equation}
We notice that $\pi^{x}_{~x}$ is the opposite of the isotropic pressure $p$. Hence, in the spacetime (2.1) one experiences different pressures on the longitudinal and transversal directions (a consequence of the direction of $v(x)$). Their signs depend on the concrete expression of the velocity field.

Keeping $v(x)$ general and using $u^{a}$ from (2.8) we find, apart from $a^{b} = 0$, that the expansion scalar $\Theta = \nabla_{a}u^{a} = v'(x)$ and
  \begin{equation}
	\sigma^{x}_{~x} = -2 \sigma^{y}_{~y} = -2 \sigma^{z}_{~z} = -\frac{\sigma^{x}_{t}}{v(x)} = \frac{2}{3} v'(x).
 \label{3.5}
 \end{equation}
In addition, $2\sigma^{2} \equiv \sigma^{ab}\sigma_{ab} = (4/3)v'^{2}$, where $\sigma^{ab}$ is the shear tensor (it is a known fact that $\sigma^{2}$ measures the local anisotropy of the spacetime \cite{WWU}).

Once the kinematical quantities have been evaluated, we are now in a position to relate the velocity field to the stress tensor. As we shell see in the next section, $\Phi(x)$ obeys an equation of motion similar to Yukawa potential one from nuclear physics. We follow up the recipe of Feng and Matzner \cite{FM} and take a hypersurface $\Sigma$ as being a level surface defined by $\Phi(x) = const.$. The normal vector field to $\Sigma$ will be given by
  \begin{equation}
	n^{a} = \frac{g^{ab} \nabla_{b} \Phi}{\sqrt{g^{ab} \nabla_{a} \Phi \nabla_{b} \Phi}},
 \label{3.6}
 \end{equation}
where the derivatives are calculated with the metric (2.1) and $\Phi(x) = -\frac{v^{2}(x)}{2}$. One finds that
  \begin{equation}
	n^{a} = \left(\frac{v}{\sqrt{1 - v^{2}}}, - \sqrt{1 - v^{2}}, 0, 0\right),
 \label{3.7}
 \end{equation}
where the x-dependence of $v(x)$ is omitted for simplicity. Note that the normal vector is spacelike, $n^{a}n_{a} = 1$ and the induced metric on $\Sigma$ is $f_{ab} = g_{ab} - n_{a}n_{b}$. To write down the surface stress tensor on $\Sigma$ \cite{EO}
  \begin{equation}
	8\pi ~\tau_{ab} = K f_{ab} - K_{ab},
 \label{3.8}
 \end{equation}
we need the components of the extrinsic curvature of $\Sigma$, defined by $K_{ab} = f^{c}_{~b} \nabla_{c} n_{a}$. For the trace $K = K^{a}_{~a}$ we get
  \begin{equation}
	K = \nabla_{a} n^{a} = \frac{1}{\sqrt{-g}} \frac{\partial}{\partial x^{a}}\left(\sqrt{-g}~n^{a}\right) = \frac{vv'}{\sqrt{1 - v^{2}}}.
 \label{3.9}
 \end{equation}
The nonzero components of $K_{ab}$ appears as
  \begin{equation}
	K_{tt} = -vv' \sqrt{1 - v^{2}},~~~K_{xx} = -\frac{v^{3}v'}{\sqrt{(1 - v^{2})^{3}}},~~~K_{xt} = -\frac{v^{2}v'}{\sqrt{1 - v^{2}}}, 
 \label{3.10}
 \end{equation}
where all quantities are evaluated on $\Sigma$.
Once (3.9) and (3.10) are plugged into (3.8), one finds that the only nonzero components are
  \begin{equation}
	8\pi ~\tau_{yy} = 8\pi ~\tau_{zz} = \frac{vv'}{\sqrt{1 - v^{2}}} = K,
 \label{3.11}
 \end{equation}
with the trace $\tau^{a}_{~a} = K/4\pi$.

 It is worth noting that our fluid is imperfect \cite{FC}. We use Barrow-Maartens prescription \cite{BM} and separate the energy density corresponding to the perfect fluid part (which is zero for our fluid) and the energy density $\rho_{a}$ corresponding to the anisotropic part, such that $\rho_{total} = \rho + \rho_{a}$, with
  \begin{equation}
	\rho_{a} = 3p = -\frac{1}{4\pi} (v'^{2} + vv'') = \frac{1}{4\pi} \frac{d^{2}\Phi}{dx^{2}}<0.
 \label{3.12}
 \end{equation}
 One observes from (3.11) that the nonzero components of $\tau_{ab}$ are those from the plane yz because the level surfaces $\Phi = const.$ means constant $v$ and $x$. If one integrates $\rho_{a}$ from some $x$ to infinity (where $v = 0)$, we obtain that integral is $\tau^{a}_{~a}$, where the approximation $v<<1$ was used, so that the square root from the denominator of (3.11) becomes unity. We also note that there is no jump of the extrinsic curvature when the surface $\Sigma$ is crossed because in either side of $\Sigma$ we have the same geometry. However, $\tau_{ab}$ appears to be nonzero thanks to the integration of $\rho_{a}$ along x-axis, as if the medium were layered transversally.

\section{Geodesics} 
In order to determine the geodesic trajectories in the geometry (2.1), we have to specify $v(x)$. Let us choose $v(x)$ such that $\pi^{x}_{~x}$ is proportional to $v^{2}(x)$. We take
   \begin{equation}
	\frac{d^{2}}{dx^{2}}\left(\frac{v^{2}(x)}{2}\right) \equiv v'^{2} + vv'' = \frac{k^{2}v^{2}}{2},
 \label{4.1}
 \end{equation}
where the constant $k$ is an inverse length, with $k = 0$ when $v$ = const. In terms of the potential $\Phi$, (4.1) may be written as
   \begin{equation}
	\Phi''(x) - k^{2} \Phi(x) = 0.
 \label{4.2}
 \end{equation}
The eq.(4.2) looks like an equation of motion for a classical massive scalar potential $\Phi(x)$. It is similar with the Yukawa potential from nuclear physics, excepting that  $\Phi(x)$ acts only along x-direction but nuclear potential is spherically symmetric. We consider $k = 2m$, where $2m$ is taken as the mass of the particle associated to the field (pion mass for nuclear forces). The eq.(4.2) may be thought of as the Poisson equation with zero mass density but adjusted with a cosmological term, as Einstein did in 1917, even though (4.2) is one dimensional.

We assume the mass $m$ is of the order of the electron mass $m_{e}$. The previous equation becomes
   \begin{equation}
	\Phi''(x) - 4m^{2} \Phi(x) = 0,
 \label{4.3}
 \end{equation}
whence $\Phi(x) = -\frac{1}{2}e^{-2mx}$, with appropriate initial conditions. Therefore, $v(x) = e^{-mx}$, because the velocity was chosen to be positive.

 The above equation (4.3) for the field $\Phi (x)$ (having units of a gravitational potential or velocity squared) may be related to the Simula \cite{TS} ''quantum pressure'' $P_{q} = -\nabla^{2}\sqrt{n}/(2m\sqrt{n})$ from his model about (2+1)-dimensional superfluid universe, where $n$ is the background condensate particle density and his $m$ is the mass of the particle. One observes that actually $P_{q}$ has units of energy and represents the so-called quantum potential $Q$ from the de Broglie-Bohm (dBB) theory \cite{RGJ, HC3}. If we take the particle density $n$ depending only on variable $x$ (a stationary fluid flow) and choosing for $P_{q}$ a constant value, say $P_{q} = -2m$, the equation for $n(x)$ becomes mathematically identical with Eq.(4.3) and so the field $\Phi(x)$ could be interpreted as being proportional to the square root of the particle density for our imperfect fluid. Let us note that Simula \cite{TS} consider  $P_{q}$ as giving rise to a gravitational field through a local density gradient and the condensate is identified as Dark Matter. We also stress that the analogy with Simula's paper works only at zero temperature when the condensate ground state is smooth and the universe is composed of Dark Matter only \cite{TS}.

\textbf{Timelike geodesics}

We have seen that the velocity 4-vector $u^{a}$ is tangent to the timelike geodesics with a null 4-acceleration, so that
 \begin{equation}
	 u^{x} = \frac{dx}{dt} = e^{-mx},
 \label{4.4}
 \end{equation}
which gives us
 \begin{equation}
x(t) = \frac{1}{m}~ ln(1 + mt),
 \label{4.5}
 \end{equation}
with $x(0) = 0$. When the fundamental constants are introduced in (4.5), one obtains for the geodesic trajectory
 \begin{equation}
x(t) = \frac{\hbar}{mc} ~ln\left(1 + \frac{mc^{2}}{\hbar}t\right).
 \label{4.6}
 \end{equation}
Our choice was $m = m_{e} = 9 .10^{-28}gm$. Hence, $\hbar/mc \approx 10^{-11} cm$ and $mc^{2}/\hbar \approx 10^{21} s^{-1}$. For example, after $t_{1} = 10^{-21}s, x(t_{1})$ takes value $(\hbar/mc) ln2 \approx 0.7 \times 10^{-11} cm$. But after $t_{2} = 1s, x(t_{2}) \approx 21\times 10^{-11} ln10 \approx 4.8 \times 10^{-10} cm$. Even for $t_{3} = 10^{18}s, x(t_{3})$ remains microscopic: $x(t_{3}) = 9\times 10^{-10} cm$. Although the velocity 
 \begin{equation}
v(t) = \frac{dx}{dt} = \frac{1}{1 + mt}
 \label{4.7}
 \end{equation}
is unity at $t = 0$ (we impose $t\geq 0$ for to get $v(t)\leq 1$); the test particle is slowing down very fast, due to the huge pressure $\pi^{x}_{~x}$ (as we shell see) and so $x(t)$ preserves its microscopic value. The fact that the instantaneous velocity of the test particle equals the speed of light in vacuo remind us the ''zitterbewegung'' phenomenon, well-known from Quantum Mechanics.
In addition, one notices that $x \rightarrow t$ when $m \rightarrow 0$ and $x \rightarrow 0$ if $\hbar \rightarrow 0$ (because $v \rightarrow 0$, too). In other words, the potential $\Phi$ is semiclassical: it depends on $\hbar$ and vanishes when $\hbar \rightarrow 0$. As a function of time, $\Phi (t) = -1/2(1+mt)^{2}$. One notices that $|\Phi|\approx 1/2$ for $t<<10^{-21}s$ and $|\Phi|<<1$ for $t>>10^{-21}s$, so that it practically vanishes for macroscopic time intervals. 

From (4.7) one can compute the acceleration $a(x) = -d\Phi/dx$, or
 \begin{equation}
a(t) = \frac{dv(t)}{dt} = -\frac{m}{(1 + mt)^{2}}.
 \label{4.8}
 \end{equation}
Numerically, the initial acceleration is $a(0) = -mc^{3}/\hbar = -3\times 10^{31} cm/s^{2}$, an extremely high (negative) value. However, after $t_{1} = 1s$, the acceleration is already 
$a(t_{1}) = -10^{-11} cm/s^{2}$. 

We now estimate the anisotropic pressures; say, along the x-direction. We have
 \begin{equation}
\pi^{x}_{~x} = \frac{1}{6\pi} \frac{m^{2}}{(1 + mt)^{2}}.
 \label{4.9}
 \end{equation}
Its initial value is $\pi^{x}_{~x}(0) = m^{2}/6\pi \approx 10^{68} N/m^{2}$, with $x(0) = 0$ and $v(0) = 1$. However, for $t >>10^{-21}s$, we might neglect the unity at the denominator and one obtains much smaller values: $\pi^{x}_{~x} \approx c^{2}/6\pi Gt^{2}$, which no longer depends on $\hbar$ or $m$. The last expression is, if we take $t \approx 10^{18}s$, the cosmological pressure, well-known from the FRW cosmology. 

\textbf{Null geodesics} 

The condition $ds^{2} = 0$ in (2.1) leads to (see also \cite{BLV})
 \begin{equation}
\left(\frac{dx}{dt}\right)^{2} - 2v(x)\frac{dx}{dt} +v^{2}(x) - 1 = 0,
 \label{4.10}
 \end{equation}
whence
 \begin{equation}
\frac{dx}{dt} = v(x) \pm 1.
 \label{4.11}
 \end{equation}
Since $v(x)$ is taken to be positive, only the minus sign in (4.11) is convenient, for to have $|dx/dt|<1$. Keeping in mind that $v(x) = e^{-mx}$, we get from (4.11)
 \begin{equation}
U(x) \equiv \frac{dx}{dt} = e^{-mx} - 1,
 \label{4.12}
 \end{equation}
which gives us
 \begin{equation}
x(t) = \frac{1}{m} ln\left[1 + e^{-mt}\right],
 \label{4.13}
 \end{equation}
with $x(0) = (1/m)ln2$. We notice in (4.12) that $U(x)\leq 0$, a consequence of having $g_{tx}<0$ in (2.1). As a function of time, $U(x)$ appears as
 \begin{equation}
U(t) = - \frac{1}{e^{mt} + 1}.
 \label{4.14}
 \end{equation}
It is worth observing that $|U(t)|<1$, as expected. Contrary to the timelike case, here we may take into account $t<0$, too. If $t \rightarrow -\infty$ (or, more physically, when $|t|>>(1/m) \approx 10^{-21}s$ for $m = m_{e}$), $U(t)$ equals -1, the velocity of light in Minkowski space. That is a consequence of the fact that the scalar potential, as a function of time, looks like
 \begin{equation}
\Phi (t) = -\frac{1}{2} e^{-2mx(t)} = -\frac{1}{2(1 + e^{-mt})^{2}}.
 \label{4.15}
 \end{equation}
Hence, $\Phi (t) \rightarrow 0$ when $t \rightarrow -\infty$ and, as a result, $x(t)$ approaches its asymptote $x(t) = -t$. In contrast, when $t \rightarrow \infty$ (or, approximating, $t>>1/m$), the velocity $U\rightarrow 0$. For example, if $t  = -10^{-15}s, U \approx -1$, and if $t = 10^{-15}s, U \approx 0$. In other words, because of the scalar field $\Phi$, the null particle is damping very fast, such that after $t = 10^{-15}s$ its velocity $U$ practically vanishes. Our interpretation here is that the time $t$ represents the duration of the measurement being performed. 

\section{Conclusions}
On the grounds of previous papers on moving fluids, we focused in this letter on the source stress tensor of the spacetime (2.1), its properties and other quantities related to it. The accelerating imperfect fluid has vanishing isotropic energy density but nonzero anisotropic stresses (positive along the direction of motion and negative isotropic pressure). The pressures depend on $\hbar$ only for time intervals much less than $1/m$. Otherwise they appear classical. We also found the timelike and null geodesics trajectories and investigated their properties.

\end{document}